[1994/12/01]

\documentclass[colordvi]{amsart}

\newif\ifpdf\ifx\pdfoutput\undefined\pdffalse\else\pdfoutput=1\pdftrue\fi

\usepackage{amscd}
\usepackage{amsthm}   
\usepackage{amsfonts}
\usepackage{amssymb}
\usepackage{amsmath}
\usepackage{verbatim}
\usepackage{alltt}    
\usepackage{color}

\ifpdf
\usepackage[pdftex,colorlinks,citecolor=blue]{hyperref} 
\else
\usepackage[hypertex,citecolor=blue]{hyperref} 
\fi

\newcommand{\ket}[1]{\ensuremath{|#1\rangle}}
\newcommand{\bra}[1]{\ensuremath{\langle #1 |}}

\newcommand{\brakett}[3]{\ensuremath{\langle #1 |#2| #3 \rangle}}

\newcommand{\TODO}[1]{} 

\newcommand{\seqA}{\ensuremath{S_1}}
\newcommand{\seqB}{\ensuremath{S_2}}
\newcommand{\seqC}{\ensuremath{S_3}}

\newtheorem{thm}{Theorem}
\newtheorem{theorem}[thm]{Theorem}
\newtheorem{defn}[thm]{Definition}
\newtheorem{lemma}[thm]{Lemma}
\newtheorem{cor}[thm]{Corollary}
\newtheorem{note}[thm]{Note}

\begin{document}

\title[Quantum Convolution Impossibility]{Quantum convolution and quantum correlation algorithms are physically impossible}

\author{Chris Lomont}
\thanks{Research supported by AFRL grant F30602-03-C-0064}

\email{clomont@cybernet.com, clomont@math.purdue.edu}

\urladdr{www.math.purdue.edu/$\tilde{~}$clomont}


\curraddr{Cybernet Systems Corporation\\
727 Airport Blvd.\\
Ann Arbor, MI, 48108-1639 USA.}

\date{Aug 20, 2003}



\begin{abstract}
The key step in classical convolution and correlation algorithms,
the componentwise multiplication of vectors after initial Fourier
Transforms, is shown to be physically impossible to do on quantum
states. Then this is used to show that computing the convolution
or correlation of quantum state coefficients violates quantum
mechanics, making convolution and correlation of quantum
coefficients physically impossible.
\end{abstract}

\maketitle

\section{Introduction}\label{s:intro}
The Fast Fourier Transform (FFT) is arguably the most important
algorithm in computer science. Many applications, from image
processing, signal processing, pattern matching, polynomial
multiplication, number multiplication, and many others are
accomplished efficiently by utilizing the FFT to compute a
Discrete Fourier Transform (DFT) of some set of data (see
\cite[Chapter 32]{Cormen01} and \cite[Chapter 3]{Gonzalez93}).
Naively, the DFT of $N=2^n$ complex values has complexity
$O(N^2)$, but the famous paper by Cooley and Tukey \cite{Cooley65}
introduced the Fast Fourier Transform, reducing the complexity to
$\Theta(N\log N)=\Theta(n2^n)$, making Fourier transforms
extremely useful in computer algorithms. This efficiency is the
basis of fast convolution and correlation algorithms, which do a
DFT on each input sequence, then a componentwise multiplication,
then an inverse DFT, computing convolution or correlation with
complexity $O(N\log N)$.

In the last decade, quantum computing has become well known due to
the integer factoring algorithm of Shor \cite{Shor94} and the
database search of Grover \cite{Grover96}, both of which have
complexities much better than their classical counterparts. There
are other problems \cite{simon94power} for which quantum computers
perform exponentially better than any classical (Turing) computer.
A quantum Fourier Transform (QFT) can be done on a quantum state
consisting of $N=2^n$ complex values with complexity $O(\log^2
N)=O(n^2)$ (see for example, \cite[Chapter 5]{ChuangNielsen}),
making it exponentially faster than the classical counterpart, but
unfortunately the peculiarities of quantum mechanics disallows
using this algorithm as a direct replacement for all FFT
algorithms. In fact, after the breakthrough algorithms of Shor and
Grover, no algorithms of similar importance have been found,
although there has been intense work in this direction.

Since the QFT and inverse QFT are more efficient than their
classical counterparts, and the FFT and inverse FFT are the
cornerstones for convolution and correlation algorithms, it is
reasonable to attempt to construct quantum analogues of
convolution and correlation algorithms that outperform their
classical counterparts. \emph{The key point of this paper is that
there is no physically realizable method to compute the normalized
convolution or correlation of the coefficients of two quantum
states.} Thus replacing classical algorithms that rely on
convolution or correlation cannot be done in a simple, direct
algorithm replacement manner, but must be approached by more
sophisticated techniques, avoiding the method examined in this
paper. The precise definitions of quantum convolution and quantum
correlation are in section \ref{s:problem}, and it is shown in
section \ref{s:impossible} that they cannot be computed without
violating quantum mechanics.

\section{Background}\label{s:background}
Since this paper touches upon several areas, we review some key
ideas to make it clearer for readers coming from different
backgrounds. In particular, we will recall how convolution and
correlation are done classically, followed by a review of rules
for quantum computation, sufficiently covered for our needs. Then
we will state the convolution and correlation problem for quantum
states. Finally, we will prove such an approach is doomed to fail,
since it would allow a violation of quantum mechanics.

\subsection{Classical convolution and
correlation}\label{s:classical}Convolution and correlation have
many uses in algorithms, which we will not go into here. Since
different authors and fields of study use slightly different
definitions for the following terms, we define them here, and
choose a form amenable to making quantum versions. In particular,
different authors use different factors where the $\sqrt{N}$
appears in our definitions, depending on how they will use them.
We choose the conventions below to make our definitions mimic the
quantum versions in section \ref{s:QFT}, where the factors are
dictated by unitarity.

\begin{defn}[Discrete Fourier Transform]
Given a sequence $\alpha_0,\alpha_1,\dots,\alpha_{N-1}$ of $N$
complex numbers, the \emph{\textbf{Discrete Fourier Transform}}
(DFT) of the sequence is defined to be the sequence
$a_0,a_1,\dots,a_{N-1}$ given by
\begin{eqnarray*}
a_j & = & \frac{1}{\sqrt{N}}\sum_{k=0}^{N-1}\alpha_k e^{2\pi i j k
/ N} \;\;\; j=0,1,\dots,N-1
\end{eqnarray*}

\end{defn}

\begin{defn}[Inverse DFT]
Given a sequence $a_0,a_1,\dots,a_{N-1}$ of $N$ complex numbers,
the \emph{\textbf{Inverse Discrete Fourier Transform}} (IDFT) of
the sequence is defined to be the sequence
$\alpha_0,\alpha_1,\dots,\alpha_{N-1}$ given by
\begin{eqnarray*}
\alpha_k & = & \frac{1}{\sqrt{N}}\sum_{j=0}^{N-1}a_j e^{-2\pi i j
k / N} \;\;\; k=0,1,\dots,N-1
\end{eqnarray*}
\end{defn}

The IDFT and DFT are inverses, so are 1-1 on sequences.

\begin{defn}[Convolution]\label{d:convolution}
The \textbf{\emph{convolution}} of two sequences of $N$ complex
numbers $\seqA = (\alpha_0,\alpha_1,\dots\alpha_{N-1})$ and $\seqB
= (\beta_0,\beta_1,\dots\beta_{N-1})$ is defined to be the
sequence $\seqC = (\gamma_0,\gamma_1,\dots,\gamma_{N-1})$ given by
$$
\gamma_k=\sum_{j=0}^{N-1} \alpha_j\beta_{k-j}\;\;\text{ for }
\;\;k=0,1,\dots N-1
$$
where subscripts are taken $\pmod{N}$.
\end{defn}

\begin{defn}[Correlation]
The \textbf{\emph{correlation}} of two sequences of $N$ complex
numbers $\seqA = (\alpha_0,\alpha_1,\dots\alpha_{N-1})$ and $\seqB
= (\beta_0,\beta_1,\dots\beta_{N-1})$ is defined to be the
sequence $\seqC = (\gamma_0,\gamma_1,\dots,\gamma_{N-1})$ given by
$$
\gamma_k=\sum_{j=0}^{N-1} \alpha^*_j\beta_{k+j}\;\;\text{ for }
\;\;k=0,1,\dots N-1
$$
where $\alpha^*_j$ is the complex conjugate of $\alpha_j$ and
subscripts are taken $\pmod{N}$.
\end{defn}

\subsubsection{Convolution/correlation algorithm}\label{s:convalgor}
It is a simple exercise to check that an algorithm to compute the
convolution (or correlation) of two length $N$ sequences in time
$O(N\log N)$ is:
\begin{enumerate}
\item FFT - apply the FFT to each sequence in time $O(N \log N)$,
giving two sequences of length $N$.

\item \label{i:multiply} Multiply - multiply the resulting
sequences componentwise in time $O(N)$, assuming each entry can be
multiplied in constant time. If doing correlation, conjugate the
first FFT sequence in time $O(N)$.

\item Invert - Apply the inverse FFT to the resulting sequence
with time complexity $O(N\log N)$, and multiply each number by
$\sqrt{N}$, giving the convolution sequence.

\end{enumerate}

\subsection{The rules of quantum computing}
A quantum system is determined by a quantum state vector, often
denoted \ket{\psi}. Mathematically, a quantum state on $n$
two-state particles (each called a qubit) is a unit vector in a
$N=2^n$ dimensional complex Hilbert space. For the rest of this
paper $N$ will be $2^n$ for some positive integer $n$. The Hilbert
space is equipped with an orthonormal basis labelled \ket{i}, for
$i=0,1,\dots,N-1$. This state is written as
$$\ket{\psi}=\sum_{i=0}^{N-1}a_i\ket{i}$$
where the $a_i$ are complex numbers, \ket{i} is the $i^{th}$
orthonormal basis element, and the $a_i$ satisfy the
\emph{normalization condition} $\sum_i |a_i|^2=1$. Quantum
mechanics allows two methods to change the state of a system:
\begin{enumerate}
\item \textbf{Unitary transformations}. Any state change of an
isolated system must be reversible, and must satisfy the
normalization condition above, which leads to unitary operations
on the state. That is, a state \ket{\psi} can be transformed to
the state $U\ket{\psi}$ via the unitary matrix $U$. Unitary means
$U^\dag U=I=U U^\dag$ where $I$ is the identity matrix, and $\dag$
is the conjugate transpose.

\item \textbf{Measurement operators}. Applying a measurement to a
quantum state returns an ``answer" with a probability related to
the magnitude of the coefficients, and places the system into the
state whose value was returned from the measurement. Precisely, a
quantum measurement consists of a collection $\{M_m\}$ of
\emph{measurement operators}, which must satisfy $\sum_m M_m^\dag
M_m = I$ to preserve probability. When applied to a system
\ket{\psi}, with probability
$$p(m)=\brakett{\psi}{M_m^\dag M_m}{\psi}$$ the state becomes the
state
$$\ket{\psi}\xrightarrow{M_m}\frac{M_m\ket{\psi}}{\sqrt{p(m)}}$$
and outcome $m$ is observed by the measurement. See \cite[Chapter
2]{ChuangNielsen} for details and examples.

\item \textbf{Combining states} The concatenation of two quantum
systems on $n$ and $m$ qubits with states \ket{\psi} and
\ket{\phi} respectively is the tensor product state
$\ket{\psi}\otimes\ket{\phi}$ in complex $2^{n+m}$ dimensional
space. Shorthand is
$$
\left(\sum_ia_i\ket{i}\right)\otimes\left(\sum_jb_j\ket{j}\right)
=\sum_{i,j}a_ib_j\ket{ij}
$$
\end{enumerate}

\TODO{clean up} For the purpose of this paper it is necessary to
define what we mean by ``physically realizable computation". There
are many approaches to this question, but for our purposes it is
enough to take a very general definition, under which we show
convolution and correlation to be impossible, since they remain
impossible under more confining and precise definitions. A quantum
computation can utilize extra states called ``ancillary qubits" as
working space, so we allow a finite number of extra qubits to be
used internally to the system.

\begin{defn}[Physically realizable]\label{d:physical} A mapping $P$ of quantum
states
\begin{eqnarray*}
\ket{\phi}&\xrightarrow{P}&\ket{\psi}
\end{eqnarray*}
is called \emph{\textbf{physically realizable}} if there exists a
finite fixed sequence of unitary transformations and measurement
operators and a quantum state \ket{\rho} (called the ancillary
qubits) performing the mapping. That is, for all \ket{\phi}, the
fixed sequence $S$ performs the mapping
\begin{eqnarray*}
\ket{\phi}\otimes\ket{\rho}&\xrightarrow{S}\ket{\psi}
\end{eqnarray*}
\end{defn}

Note in particular, the mapping is a composition of linear
operators, so is a linear operator. There are some researchers
working on nonlinear quantum mechanics, but there is no
experimental evidence to date that supports nonlinearity
\cite{Majumder90}. One interesting result is that if quantum
mechanics is nonlinear, then quantum computers can solve
NP-complete and \#P problems in polynomial time
\cite{AbramsLloyd98}, which some believe is very good evidence
that quantum mechanics is strictly linear. For our purposes we use
the standard view that quantum mechanics has linear evolution.

\subsubsection{The Quantum Fourier Transform}\label{s:QFT}
Similar to the classical Fourier Transform and inverse defined in
section \ref{s:classical}, we can define quantum analogues,
operating on sequences of complex numbers stored as quantum state
coefficients.
\begin{defn}[Quantum Fourier Transform]
The \emph{\textbf{Quantum Fourier Transform}} (QFT) is the unitary
map defined on basis states \ket{j} as
\begin{eqnarray*}
\ket{j} & \xrightarrow{QFT} &
\frac{1}{\sqrt{N}}\sum_{k=0}^{N-1}e^{2\pi i j k / N}\ket{k}
\end{eqnarray*}
and extended by linearity.
\end{defn}
\begin{defn}[Inverse QFT]
The \emph{\textbf{Inverse Quantum Fourier Transform}} (IQFT) is
the unitary map defined on basis states \ket{k} as
\begin{eqnarray*}
\ket{k} & \xrightarrow{IQFT}  &
\frac{1}{\sqrt{N}}\sum_{j=0}^{N-1}e^{-2\pi i j k / N}\ket{j}
\end{eqnarray*}
and extended by linearity.
\end{defn}

\begin{note}
\emph{It is easy to check these are inverses on quantum states,
and that each transform is unitary.}
\end{note}

The quantum state $\ket{\psi}=\sum a_i \ket{i}$ then transforms as
\begin{eqnarray}\label{e:QFTtransform}
\sum_{j=0}^{N-1}a_j\ket{j}&\xrightarrow{QFT}& \sum_j a_j \frac{1}{\sqrt{N}}\sum_{k=0}^{N-1}e^{2\pi i j k / N}\ket{k}\\
&=& \sum_k\left(\frac{1}{\sqrt{N}}\sum_ja_je^{2\pi i j k
/N}\right)\ket{k}
\end{eqnarray}

One reason some quantum algorithms are superior to classical ones
is that the QFT has complexity $O(\log^2 N)=O(n^2)$ \cite[Chapter
5]{ChuangNielsen}, \emph{which is exponentially faster than the
classical $O(n2^n)$ version!} It is precisely this exponential
speedup that allows Shor's integer factoring algorithm to
out-perform the best known classical one.

\section{Problem Statement}\label{s:problem}

In this section we describe an attempt to construct quantum
versions of the convolution and correlation algorithms, based on
the classical algorithms \ref{s:convalgor}. Since the QFT operates
on quantum coefficients, we will encode the sequences in a quantum
state, with the sequence entries stored in the coefficients. Since
a quantum state $\sum_i\alpha_i\ket{i}$ requires
$\sum_i|\alpha_i|^2=1$, we require each sequence to have at least
one nonzero entry, and then we normalize the sequence to have norm
1. We ignore the physical method the quantum state is constructed
from the sequence, since the many ways to construct it would lead
us. So we have a (hypothetical) method of converting a nonzero
sequence of complex numbers into a representative quantum state.

The problems become:

\begin{defn}[The Quantum Convolution Problem]\label{d:problem1}
Given quantum states representing the two initial sequences,
compute a quantum state representing the convolution sequence.
That is, given the two states (for $N=2^n$)
\begin{eqnarray}
\ket{\alpha} & = & \sum_{i=0}^{N-1}\alpha_i\ket{i}\\
\ket{\beta} & = & \sum_{j=0}^{N-1}\beta_j\ket{j}
\end{eqnarray}
compute a state
\begin{equation}
\ket{\gamma}  =  \sum_{k=0}^{N-1}\gamma_k\ket{k}
\end{equation}
where $\ket{\gamma}$ represents the normalization of the sequence
given by
\begin{equation}
c_k  =  \sum_{j=0}^{N-1}\alpha_j\beta_{k-j}\;\;\text{ for
}k=0,1,\dots N-1
\end{equation}
and subscripts are taken $\pmod{N}$.
\end{defn}

\begin{defn}[The Quantum Correlation Problem]\label{d:problem2}
Given quantum states representing the two initial sequences,
compute a quantum state representing the correlation sequence.
That is, given the two states (for $N=2^n$)
\begin{eqnarray}
\ket{\alpha} & = & \sum_{i=0}^{N-1}\alpha_i\ket{i}\\
\ket{\beta} & = & \sum_{j=0}^{N-1}\beta_j\ket{j}
\end{eqnarray}
compute a state
\begin{equation}
\ket{\gamma}  =  \sum_{k=0}^{N-1}\gamma_k\ket{k}
\end{equation}
where $\ket{\gamma}$ represents the normalization of the sequence
given by
\begin{equation}
c_k  =  \sum_{j=0}^{N-1}\alpha^*_j\beta_{k+j}\;\;\text{ for
}k=0,1,\dots N-1
\end{equation}
and subscripts are taken $\pmod{N}$.
\end{defn}

\begin{note}\label{n:padding}
\emph{For each definition to make sense, the resulting sequence
$c_k$ cannot be all zeros. This can be guaranteed by requiring
that the initial sequences are each not all zero, and then padding
each initial sequence by appending $N$ zeros. Then, choose $i_0$
and $j_0$ each minimal so that $\alpha_{i_0}\neq 0$ and
$\beta_{j_0}\neq 0$. Then $c_{i_0+j_0}=\alpha_{i_0}\beta_{j_0}\neq
0$ in the convolution sequence. Pick $i_1$ maximal so that
$\alpha_{i_1}\neq 0$, then
$c_{j_0-i_1}=\alpha^*_{i_1}\beta_{j_0}\neq 0$ in the correlation
sequence, with subscripts taken $\pmod{N}$. Padding by $N$ zeros
requires adding only one more qubit, so we assume for the rest of
this paper that this is the case. Then each problem is
well-defined. This changes each problem slightly, but proving
impossibility in these slightly amended case suffices to prove
impossibility in general.}
\end{note}

\section{Impossibility proof}\label{s:impossible}
This section gives a proof that the problems given in definitions
\ref{d:problem1} and \ref{d:problem2} cannot be computed by any
device obeying quantum mechanics. Precisely, we prove in section
\ref{s:final}:

\begin{theorem}[Impossibility of quantum convolution]\label{t:impossible1}There is no physically realizable process $P$ to
compute the (normalized) convolution of the coefficients of two
quantum states. That is, for arbitrary quantum states
$\sum_ia_i\ket{i}$ and $\sum_jb_j\ket{j}$, there is no physically
realizable process $P$ to compute the state
\begin{equation}
\sum_{i,j=0}^{N-1}a_ib_j\ket{ij}\xrightarrow{P}\lambda\sum_{k=0}^{N-1}\sum_{j=0}^{N-1}a_jb_{k-j}\ket{k}
\end{equation}
where $\lambda=1/\sqrt{\sum|a_ib_j|^2}$ is the normalization
factor, subscripts are taken $\pmod{N}$, and $N=2^n$ for some
integer $n>0$.
\end{theorem}

and

\begin{theorem}[Impossibility of quantum correlation]\label{t:impossible2}There is no physically realizable process $P$ to
compute the (normalized) correlation of the coefficients of two
quantum states. That is, for arbitrary quantum states
$\sum_ia_i\ket{i}$ and $\sum_jb_j\ket{j}$, there is no physically
realizable process $P$ to compute the state
\begin{equation}
\sum_{i,j=0}^{N-1}a_ib_j\ket{ij}\xrightarrow{P}\lambda\sum_{k=0}^{N-1}\sum_{j=0}^{N-1}a^*_jb_{k+j}\ket{k}
\end{equation}
where $\lambda=1/\sqrt{\sum|a_ib_j|^2}$ is the normalization
factor, subscripts are taken $\pmod{N}$, and $N=2^n$ for some
integer $n>0$.
\end{theorem}

This is done in several parts. First, we show that step
\ref{i:multiply} in algorithm \ref{s:convalgor} has no quantum
analogue by studying requirements on a linear transformation that
attempts step \ref{i:multiply}. Then we show no physical process
consisting of arbitrary sequences of unitary operations and
measurements can perform step \ref{i:multiply}. This proves that
step \ref{i:multiply} is impossible to compute on quantum states.
Finally, we show that any physical process able to compute quantum
convolution or correlation as in definitions \ref{d:problem1} and
\ref{d:problem2} would be able to compute the impossible step
\ref{i:multiply}, a contradiction. Thus there can be no quantum
convolution or correlation done on quantum states.

\subsection{Linearity considerations}\label{s:linearity}
Suppose we try to obtain the componentwise product of two states,
using some linear operator on the initial state. Heuristically,
this should fail, since all components in the outcome may be zero.
Even if we remove this case (which is reasonable, see note
\ref{n:padding}), we show such an operation is still not possible.
Since linear operators encompass both unitary operations and
measurement systems, we analyze the linear operator case first. As
in definition \ref{d:physical}, we allow the algorithm to use
arbitrary finite ``workspace" in the form of a third quantum state
$\sum_kc_k\ket{k}$ on $m$ qubits in order to show the problem
cannot even be computed with extra working space.

\begin{lemma}\label{l:linearity1}
There is no linear operator $L$ and a quantum state $\sum
c_k\ket{k}$ such that for arbitrary quantum states $\sum
a_i\ket{i}$ and $\sum b_j\ket{j}$ with $\sum|a_ib_i|\neq 0$, the
following operation is performed on quantum states:
\begin{equation}
\sum_{i,j,k}a_ib_jc_k\ket{ijk}\xrightarrow{L}\left(\lambda\sum_i
a_ib_i\ket{i}\right)\otimes\left(\sum_{j,k}d_{jk}\ket{jk}\right)\label{e:linearity1}
\end{equation}
where the $d_{jk}$ are functions of the $a_i, b_j$, and $c_k$,
with $\sum |d_{jk}|^2=1$, and $\lambda=1/\sqrt{\sum_i|a_ib_i|^2}$
is a normalization factor.
\end{lemma}
\begin{proof}
Assume there is a linear operator $L$ and initial state
$\sum_kc_k\ket{k}$ satisfying \ref{e:linearity1}. Write $L$ as
$$L=\sum_{\substack{a,b,c\\r,s,t}}e_{rstabc}\ket{rst}\bra{abc}$$
and apply it to the initial state to get
\begin{eqnarray}
\left(\sum_{\substack{a,b,c\\r,s,t}}e_{rstabc}\ket{rst}\bra{abc}\right)\left(\sum_{i,j,k}a_ib_jc_k\ket{ijk}\right)&=&\sum_{\substack{r,s,t\\i,j,k}}a_ib_jc_ke_{rstijk}\ket{rst}
\end{eqnarray}
We want this state to equal the desired outcome, giving
\begin{equation}
\sum_{\substack{r,s,t\\i,j,k}}a_ib_jc_ke_{rstijk}\ket{rst} =
\lambda\sum_{r,s,t}a_rb_rd_{st}\ket{rst}
\end{equation}
Looking at the \ket{0st} component, we require for a fixed $s, t$
that
\begin{equation}\label{e:rel1}
\lambda a_0b_0d_{st}=\sum_{i,j,k}a_ib_jc_ke_{0stijk}
\end{equation}
Equation \ref{e:rel1} must hold for any $a_i$ and $b_j$
coefficients giving a quantum state. Fix $\epsilon\in(0,1)$, and
pick sequences of coefficients as
\begin{eqnarray}
\{a_i\} &=& \{\epsilon,\sqrt{1-\epsilon^2},0,0,0\dots,0\}\\
\{b_j\} &=& \{1,0,0,\dots,0\}
\end{eqnarray}
Then $\lambda = 1/\epsilon$, $a_0b_0=\epsilon$,
$a_1b_0=\sqrt{1-\epsilon^2}$, and $a_ib_j=0$ for all other $i,j$
combinations. Equation \ref{e:rel1} becomes
\begin{eqnarray}
d_{st} &=& \epsilon\sum_k c_k e_{0st00k}+\sqrt{1-\epsilon^2}\sum_k c_k e_{0st01k}\\
&=& \epsilon C_1 + \sqrt{1-\epsilon^2}C_2
\end{eqnarray}
where $C_1$ and $C_2$ are constants depending on the initial
choices for $L$ and $c_k$. Normalization requires
\begin{eqnarray}
1 &=& \sum_{s,t}|d_{s,t}|^2\\
 &=& \sum_{s,t}|\epsilon C_1 + \sqrt{1-\epsilon^2}C_2|^2\\
 &=& MN\left(\epsilon^2 C_1^2 +
 2\epsilon\sqrt{1-\epsilon^2}C_1C_2+(1-\epsilon^2)C_2^2\right)\\
 &=& MNC_2^2 + \epsilon^2MN(C_1^2-C_2^2)+2\epsilon\sqrt{1-\epsilon^2}MNC_1C_2\label{e:contradiction}
\end{eqnarray}
Equation \ref{e:contradiction} must hold true for all
$\epsilon\in[0,1]$. When $\epsilon=0$, the last two terms are
zero, forcing $C_2=1/\sqrt{MN}\neq 0$. When $\epsilon=1$, the last
term is zero, and the middle term must be zero, forcing
$C_2^2=C_1^2\neq 0$. Now the middle term is zero for all
$\epsilon$. However, when $\epsilon=1/2$, equation
\ref{e:contradiction} is a contradiction, so there can be no
linear operator $L$ and initial state $\sum c_k\ket{k}$ satisfying
\ref{e:linearity1} for arbitrary input states.

\end{proof}
Note lemma \ref{l:linearity1} is a purely mathematical statement,
driven by physical requirements, and cast in quantum mechanics
language. The following variation is used later to prove the
impossibility of quantum correlation.
\begin{cor}\label{l:linearity2} There is no linear operator $L$ and a quantum
state $\sum c_k\ket{k}$ such that for arbitrary quantum states
$\sum a_i\ket{i}$ and $\sum b_j\ket{j}$ with $\sum|a_ib_i|\neq 0$,
the following operation is performed on quantum states:
\begin{equation}
\sum_{i,j,k}a_ib_jc_k\ket{ijk}\xrightarrow{L}\left(\lambda\sum_i
a^*_ib_i\ket{i}\right)\otimes\left(\sum_{j,k}d_{jk}\ket{jk}\right)\label{e:linearity2}
\end{equation}
with $\sum |d_{jk}|^2=1$, and $\lambda=1/\sqrt{\sum_i|a_ib_i|^2}$
is a normalization factor.
\end{cor}
\begin{proof}
The proof is the same as lemma \ref{l:linearity1}, since the
sequences needed in the proof were real numbers.
\end{proof}

\subsection{Unitary transforms and measurements fail}\label{s:fail1}
\begin{thm}[]\label{t:notransform}
There is no physical process $P$ capable of either of the
transformations
\begin{eqnarray}
\sum_{i,j}a_ib_j\ket{ij}&\xrightarrow{P}&\lambda\sum_ia_ib_i\ket{i}\label{e:notrans1}\\
\sum_{i,j}a_ib_j\ket{ij}&\xrightarrow{P}&\lambda\sum_ia^*_ib_i\ket{i}\label{e:notrans2}
\end{eqnarray}
on arbitrary quantum states with $\sum|a_ib_i|\neq 0$, where
$\lambda$ is a normalization factor.
\end{thm}
\begin{proof}
Since unitary transforms are linear operators, both lemma
\ref{l:linearity1} and corollary \ref{l:linearity2} remain true
when the phrase ``linear operator $L$" is replaced with ``unitary
transform $L$". So no single unitary operation $L$ can perform
either transformation. Similarly, a measurement operator $\{M_m\}$
is a collection of linear transforms on a quantum state, so no
single measurement operator can perform either transformation.
Finally, given any (finite) sequence of unitary operations and
measurement operations, we can always apply all the unitary
operations first, and then apply the measurements, by the
``Principle of Deferred Measurement" \cite[Chapter
4]{ChuangNielsen}. Then the unitary operations compose to a single
unitary operation, and the measurements combine to form a single
measurement operation \cite[Exercise 2.57]{ChuangNielsen}. A
unitary operation $U$ followed by a measurement operator $\{M_m\}$
is the measurement operator $\{UM_m\}$, which also cannot perform
either transformation. So there is no physical process $P$ to
perform either transform.
\end{proof}

\subsection{Quantum convolution and correlation are impossible}\label{s:final}
Finally, we can prove the main theorems \ref{t:impossible1} and
\ref{t:impossible2}:

\begin{proof}[Proof of theorem \ref{t:impossible1}]
Suppose such a process $P$ existed. Then the composition
$\left(IQFT\otimes IQFT\right)\circ P \circ QFT$ is physically
realizable, which performs transformation \ref{e:notrans1} in
theorem \ref{t:notransform}, a contradiction.

\TODO{diagram removed here}
Thus there can be no physically realizable quantum convolution
process.
\end{proof}

Note: this proof could use the fact that $P$ would be linear, as
would the composition, which would contradict lemma
\ref{l:linearity1}.

\begin{proof}[Proof of theorem \ref{t:impossible2}]
The proof is the same as the proof of theorem \ref{t:impossible1}
with minor changes. Thus there can be no physically realizable
quantum correlation process.
\end{proof}

\section{Conclusion and open problems}
We have shown there is no physical way to compute the convolution
or correlation on quantum states. This is unfortunate for many
applications, yet there may remain ways to attack specific
sequences which yield efficient quantum algorithms. For example,
it is possible to use a measurement to compute the transform $\sum
a_ib_j\ket{ij}\rightarrow\lambda\sum a_ib_i\ket{ii}$, but the
probability of obtaining this state tends to 0 as
$N\rightarrow\infty$, likely rendering it useless in algorithms.

A final interesting point is that given a sequence of numbers on a
classical computer, we could compute a convolution, and prepare
the quantum states in lemma \ref{l:linearity1}, which seems to
violate theorem \ref{t:notransform}. The point is that once the
input sequences are in the quantum states, without knowing the
classical information so we can make copies, there is no way to
get the output state. Quantum states cannot be copied by the No
Cloning Theorem \cite{Dieks82,WZ82}, otherwise the transform of
theorem \ref{t:notransform} could be computed by sampling enough
copies of the input to determine the classical sequences,
classically computing the convolution, then constructing the
output state. This approach is prevented by the No Cloning
Theorem. The key point is that storing a classical information
theory sequence of numbers into the quantum state coefficients is
not a reversible process, since there is no way to read all the
coefficients back out of the state.

A final note: this result was inspired by a comment made by David
Meyer, who obtained similar results independently.


\bibliographystyle{amsplain}
\bibliography{qbib,cbib}

\end{document}